\documentclass[reprint,amsmath,amssymb,aps]{revtex4-1}

\pdfoutput=1

\usepackage{graphicx,bm}
\usepackage{natbib}
\usepackage{pstricks}
\usepackage{times}

\newcommand{\dinv}{d^{-1}}

\newcommand\bx{\ensuremath{{\bf x}}}
\newcommand\bxi{\ensuremath{{\bf x}_i}}

\newcommand\CC{\ensuremath{\mathcal{C}}}

\begin{document}

\preprint{}

\title{A self-similar cascade of instabilities in the surface
quasigeostrophic system}

\author{R. K. Scott}
\email{rks@mcs.st-and.ac.uk}

\author{D. G. Dritschel}%

\affiliation{School of Mathematics and Statistics,
University of St Andrews, St Andrews KY16 9SS, Scotland}

\date{\today}

\sloppy

\begin{abstract}
We provide numerical evidence for the existence of a cascade of
filament instabilities in the surface quasigeostrophic system for
atmospheric and oceanic motions near a horizontal boundary.  The
cascade involves geometrically shrinking spatial and temporal scales
and implies the singular collapse of the filament width to zero in a
finite time.  The numerical method is both spatially and temporally
adaptive, permitting the accurate simulation of the evolution over an
unprecedented range of spatial scales spanning over ten orders of
magnitude.  It provides the first convincing demonstration of the
cascade, in which the large separation of scales between subsequent
instabilities has made previous numerical simulation difficult.
\end{abstract}

\maketitle

The surface quasigeostrophic system is a two-dimensional active scalar
equation developed as a model for the slow, large-scale motions of
rotating, stratified flows of planetary atmospheres and oceans near a
horizontal surface or strong jump in stratification such as the
midlatitude tropopause \citep{charney:1948,hoskins:1972,blumen:1978}.
It may be expressed by the material advection of a surface temperature
field $\theta(x,y,t)$ by a flow to which it is diagnostically related:
\begin{equation}\label{eq1}
\frac{\partial\theta}{\partial t} + J(\psi,\theta)=0
\qquad
\theta=-(-\Delta)^{1/2}\psi,
\end{equation}
where $\psi$ is the streamfunction, $\Delta$ is the two-dimensional
Laplacian, and where $J(\cdot,\cdot)$ is the Jacobean determinant
\citep{held:1995}.  It differs from the usual two-dimensional Euler
equations of incompressible fluid motion through the relation between
$\theta$ and $\psi$, where the more local relation in the surface
quasigeostrophic case gives rise to more energetic small-scale
motions.  In fact, the advection term in \eqref{eq1} possesses a
quadratic degree of nonlinearity similar to that of the
three-dimensional Euler equations, for vorticity in the case of the
Euler equations, for the skew gradient of temperature in the case of
the surface quasigeostrophic equations
\citep{constantin:1994a,constantin:1995,tran:2010}; see in particular
equation (13) of Ref.\cite{constantin:1995}.  The similarities between
\eqref{eq1} and the three-dimensional Euler equations have stimulated
much research into the regularity of the surface quasigeostrophic
system, in particular, whether a singularity in the temperature
gradient $\nabla\theta$ may form in finite time from smooth initial
conditions, or whether weak solutions may possess a finite dissipation
anomaly in the inviscid limit, the counterpart to the famous Onsager
conjecture for turbulent flow governed by the three-dimensional Euler
equations \citep{onsager:1949,ohkitani:1997}.  Despite the reduced
dimensionality, however, the presence of singularities in the surface
quasigeostrophic system remains an open problem.


When the initial distribution of $\theta$ is smooth, it is known that
a singularity may not form in finite time if the geometry of the
temperature field takes the form of a closing saddle, involving the
growth of unbounded gradients of $\theta$ along a line
\citep{cordoba:1998}.  Numerical studies of the closing saddle
\citep{ohkitani:1997,constantin:1998} indicate double exponential
growth $\nabla\theta\propto\exp\exp t$.  Singularities may yet form at
isolated points.  One scenario for this has been proposed involving a
cascade of repeated filament instabilities of geometrically shrinking
scale \citep{hoyer:1982,pierrehumbert:1994,held:1995}.  Because
filament growth rates scale with the inverse filament width, were such
a cascade of instabilities to follow a self-similar pattern then a
singularity involving the blow-up of $\nabla\theta$ would occur in
finite-time.  However, a large scale separation between subsequent
instabilities (see below) means that such a cascade has been extremely
difficult to demonstrate numerically, the most recent results being
limited to at best two or three consecutive instabilities before the
limits of numerical resolution are reached \citep{scott:2011}.
Furthermore, uncertainties remain over the role of the numerical grid
in the triggering of each instability.


In the special case where the temperature distribution comprises a
discrete patch (uniform inside a closed contour, zero outside, for
example), the relation between $\theta$ and $\psi$ may be expressed in
terms of a single contour integral around the patch boundary, in a
manner directly analogous to the Biot-Savart law relating electric
current to magnetic field.  Representing the patch boundary by a
closed contour $\CC$, its evolution is then governed by
\begin{equation} \label{evol}
\frac{d\bxi}{dt}=\frac{\theta_0}{2\pi}\int_\CC \frac {d\bx'}{|\bxi-\bx'|}
\end{equation}
for each point $\bxi$ on the contour $\CC$.  Although the
discontinuity in $\theta$ at the patch boundary implies a logarithmic
divergence of the velocity component tangential to $\CC$
\citep{juckes:1995}, this component has no effect on the boundary
evolution, which is determined entirely by the component normal to
$\CC$, and the problem is well-posed.
The regularity problem for this system is concerned with the formation
in finite time of a discontinuity in the patch boundary, either
through the development of a corner with infinite curvature at a
point, or the approach to zero of the minimum distance between two
contour segments.  In contrast to the smooth case examined in
Ref~\cite{scott:2011} the filament cross section formed by
instabilities of the patch is guaranteed to retain the same trivial
profile throughout the evolution.  Further, the reduction of the
dynamics in the patch case to the evolution of a single contour
enables the application of a highly efficient Lagrangian contour
dynamical algorithm, that is here able to capture the patch evolution
through a series of instabilities spanning a range of spatial scales
on the order of $10^{10}$.  The cascade is shown to be self-similar
upon a uniform rescaling of the independent variables.

For completeness, we note that another potential route to singularity
for the patch was identified in numerical experiments of
\citep{cordoba:2005}, involving the formation of a corner.  While
mathematically singular, it is important to recognize that this type
of singularity has no counterpart in the evolution of the continuous
$\theta$ distribution of most relevance to fluid motion: a continuous
distribution may develop a corner in a level set of $\theta$ without
any accompanying blow-up of the temperature gradient.  One reason that
the instability cascade is important is that the counterpart in the
evolution of the continuous distribution may also exist (see below).
The generalization of the present result to the more general case of
the continuous distribution will be addressed in future work.


We solve the evolution of the patch using the contour dynamics method
of Dritschel \citep{dritschel:1989}, representing the patch boundary
by a series of points along a contour that are redistributed according
to the boundary shape.  The method has been refined here in that the
integral \eqref{evol} around \CC\ is now computed along a series of
global splines connecting the points, ensuring continuity of curvature
and the tangent vector at each point.  The point distribution along
\CC\ is determined adaptively according to the curvature of \CC, the
local value of the strain field, and the minumum separation between
adjacent segments of the boundary, more points being added to maintain
accuracy as the complexity of the boundary increases.  This ensures
that accuracy is maintained even in regions of very high curvature and
where filaments become narrow.

\begin{figure}
\centerline{
\includegraphics[width=8cm]{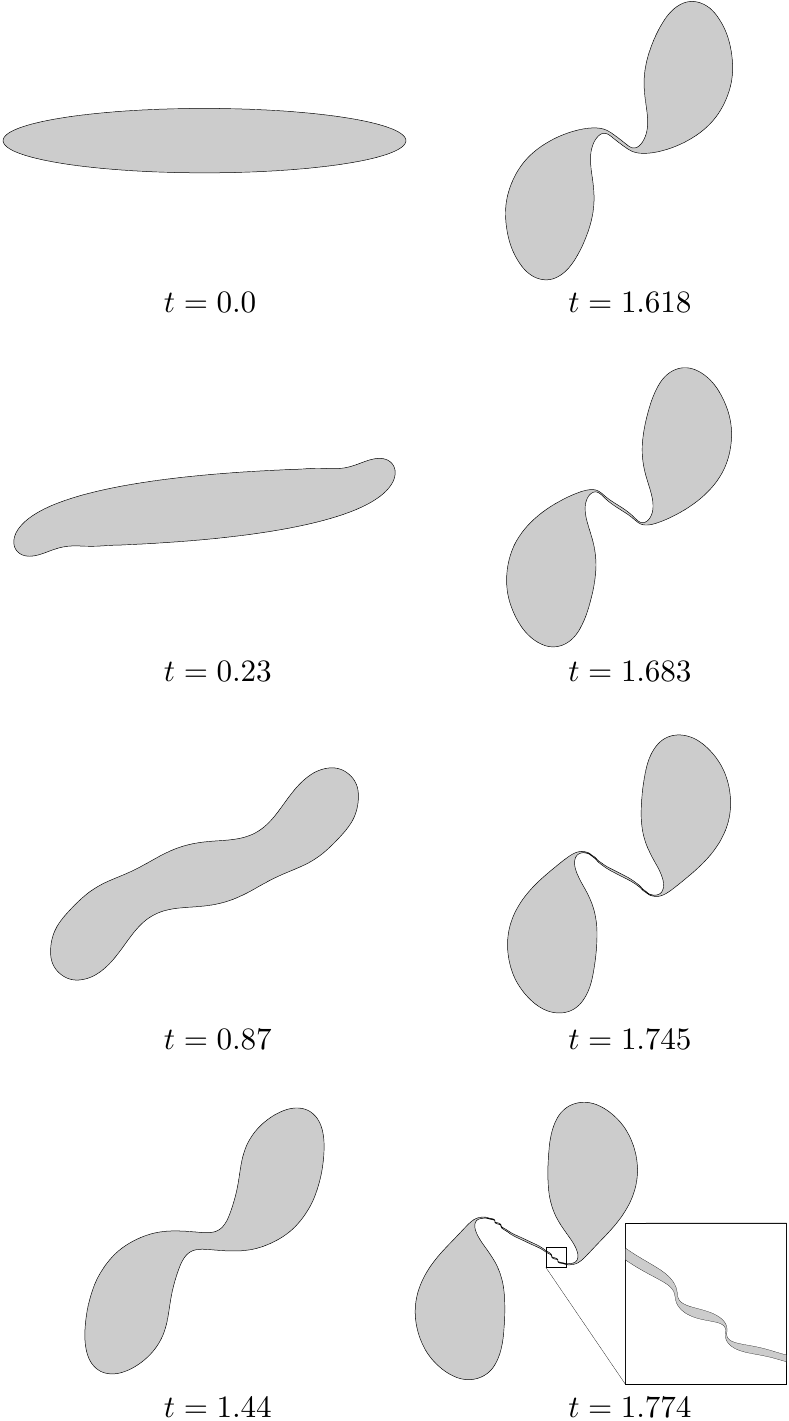}
}
\caption{Snapshots of the patch boundary at selected times prior to
the onset of the first filament instability; $\theta=2\pi$ inside the
patch boundary $\theta=0$ outside.  Final frame shows an $8\times$
magnification of a portion of the central filament.}
\label{f:patch}
\end{figure}


To illustrate the cascade, we consider a patch of $\theta$ defined by
an initial elliptical boundary of aspect ratio $a=0.16$, as shown in
the upper left panel of Figure~\ref{f:patch}, with $\theta=2\pi$
inside the boundary and $\theta=0$ outside.  As time increases the
ellipse rotates and deforms, with the tips developing into two
distinct lobes separated by a filament ($t=1.44$) that subsequently
lengthens and thins ($t=1.618$ to $t=1.745$).  By $t=1.774$ an
instability of the central filament has occurred with the development
of a series of new lobes connected by new filaments at a much smaller
scale (inset).  These filaments subsequently become unstable and role
up in a self-similar manner on a shrinking timescale (an animation of
the patch evolution is available as online supplementary material).

\begin{figure}
\centerline{
\includegraphics[width=7cm]{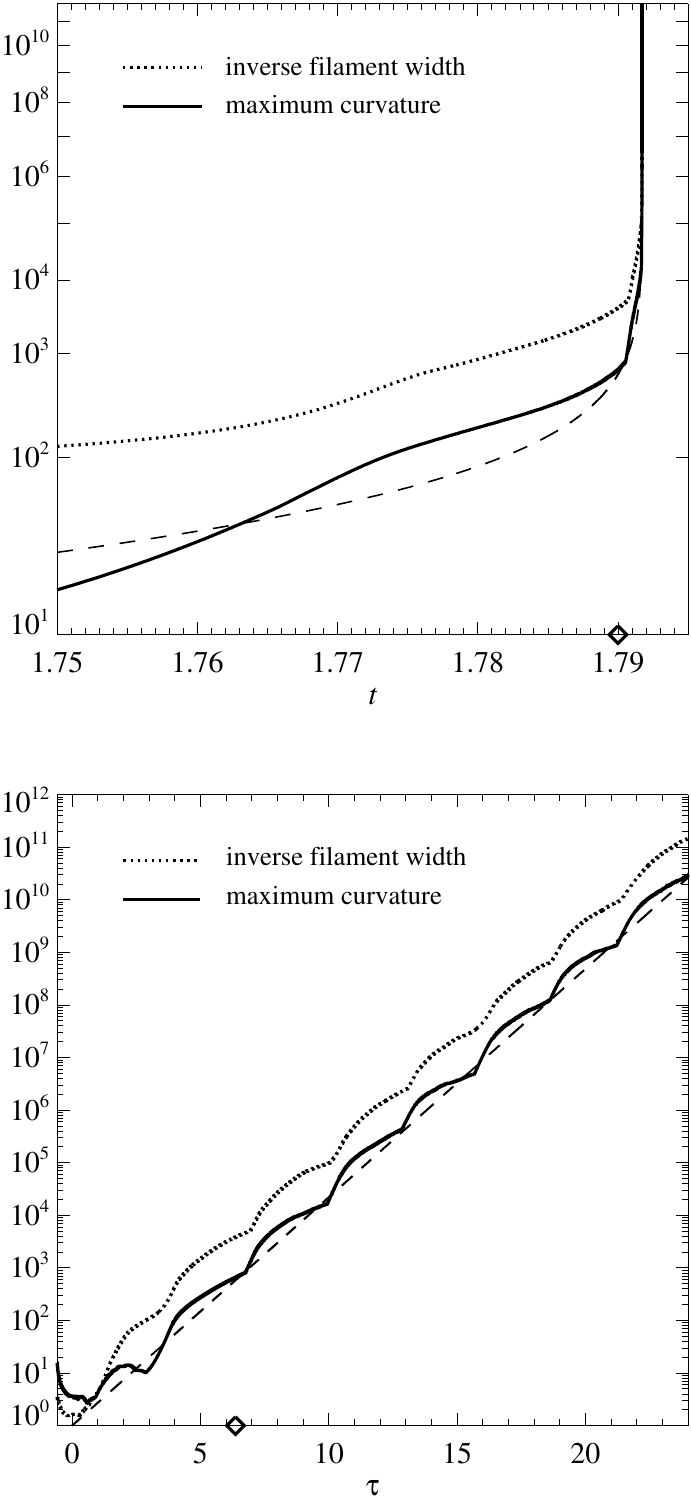}
}
\caption{Maximum curvature and inverse filament width as a function of
time $t$ (a) and rescaled time $\tau=-\log(t_s-t)$, with
$t_s=1.7917134235333878$ (b).  Corresponding times in each panel are
indicated by the diamond symbol; the dashed lines indicate the
function $1/(t_s-t)$.  Note the double-log scale in (a) and the
single-log scale in (b).}
\label{f:curv}
\end{figure}

The rapid collapse of both time and length scales can be seen by
considering the growth of curvature $\kappa$ and inverse minimum
cross-filament width $\dinv$ in time.  The development of the filament
instability between $t=1.745$ and $t=1.774$ involves the accelerating
growth in both $\kappa$ and $\dinv$, Figure~\ref{f:curv}(a).  At
around $t=1.774$ this growth slows as the new filament is more
gradually extended.  Fast growth begins again at around $t=1.791$ with
the onset of the next filament instability; the shrinking time scales
suggest unbounded growth in both $\kappa$ and $\dinv$ at a finite time
$t_s$, which is here estimated as $t_s=1.79171342353339$; both
$\kappa$ and $\dinv$ follow approximately $1/(t_s-t)$, though with
alternating periods of more or less rapid growth according to whether
the smallest filament is in a roll-up or extending phase.

The evolution as the singularity is approached is clarified by
introducing a rescaled time variable \citep{cordoba:2005}
$\tau=-\log(t_s-t)$.  The repeated stages of rapid growth (instability
development) followed by saturation stage (filament extension prior to
the onset of the next instability) occur multiple times, with the
filament scale shrinking by a factor of around twenty each time,
Figure~\ref{f:curv}(b).  At each stage the new filament becomes
unstable when it shrinks to a width of about one-twentieth that of the
parent filament, independent of the scale of the filament relative to
the original ellipse.  The time taken for the development of each
filament is also one-twentieth that of the previous one, implying a
collapse of filament width to zero in a finite time.

\begin{figure*}
\centering{
\includegraphics[width=15cm]{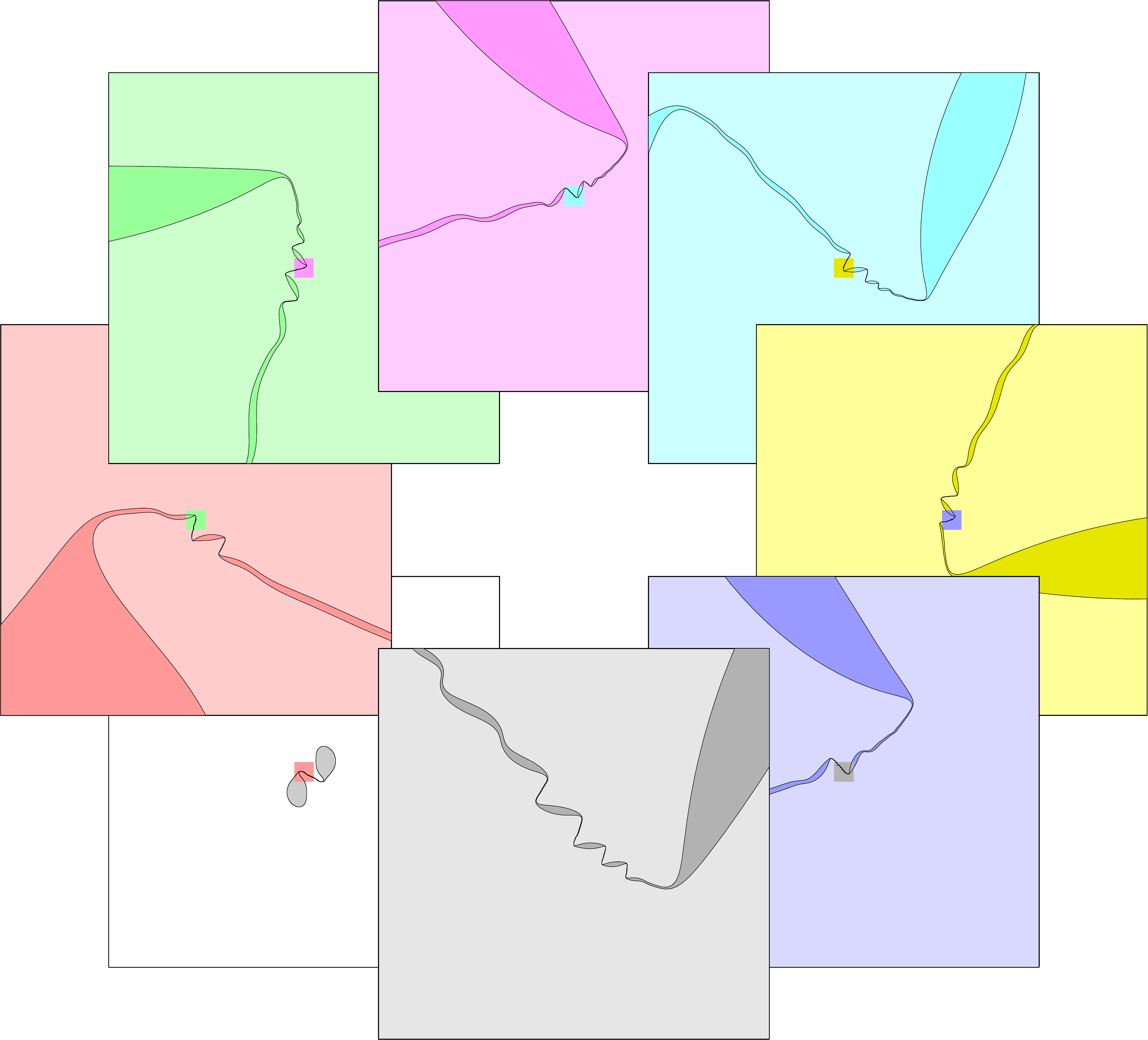}
}
\caption{Contour shape at time $t=1.79171342351654$, or
$\tau=24.8068$, at successive magnifications of $20\times$ (linear
dimension).  The central shaded square in each panel is shown
magnified in the adjacent panel of the same color.}
\label{f:zoom}
\end{figure*}

Approaching the singularity time $t_s$, the rapid acceleration of the
evolution at the smallest scales means that the large scale field
becomes effectively frozen.  The full evolution up to this time is
presented in the supplementary material, which makes a continuous
rescaling of the spatial variables by the instantaneous value of the
maximum curvature $\kappa$.  Near the final time of the integration
$t=1.79171342351654$, or $t_s-t=1.7\times 10^{-11}$, the shape of the
boundary is as shown in Figure~\ref{f:zoom}, where the series of
panels depicts successive magnifications by a linear factor of twenty
centered on the point of maximum curvature, each central shaded square
being magnified in the adjacent panel of the same color.  Despite
differences in the details the filament evolution, the large-scale
structure at each level of magnification is remarkably uniform.  Note
that each magnification appears rotated through a constant angle,
implying a uniform angular velocity in the re-scaled time $\tau$.


The physical cause and uniform scale reduction of the instability
cascade may be understood from consideration of the stability
properties of a single filament in the surface quasi-geostrophic
system.  Linear growth rates for the filament instability scale as the
inverse filament width: the inversion relation in \eqref{eq1} means
that velocity values scale as $\theta_0$, which implies that the time
taken for disturbances to grow to amplitudes comparable to the
filament width are proportional to the filament width itself.  We note
that such a cascade is typically precuded in the two-dimensional
vorticity equation because of the stabilizing effect of background
shear on the growth of filament disturbances \citep{dritschel:1991},
and because growth rates do not increase as the filament width
decreases.

The independence of the rescaling factor on scale suggests that each
instability is governed principally by the local shape of the patch in
the vicinity of the filament.  Because interactions in the
quasigeostrophic model are more local than for the case of
two-dimensional vortex dynamics, the evolution of a given filament is
to a good approximation controlled by the influence of the parent
filament that spawned it.  However, the contribution to the velocity
field by a local approximation to the contour integral in fact
diverges logarithmically, meaning that the precise evolution requires
consideration of the global patch shape.  The dynamics is thus
expected to be non-universal and may depend upon details of the
initial conditions.  Indeed, for initially elliptical patches, a
variety of different evolutions are obtained for different values of
the aspect ratio $a$. For $a$ greater than about $0.19$ the evolution
appears to remain smooth for all time; for $a$ in the range
$(0.17,0.19)$ the patch boundary develops a corner before disturbances
of the central filament have time to grow to large amplitude; for $a$
below about $0.12$ a more complex combination of corner formation and
filament instability occurs.  A full description will be reported in a
subsequent paper.

While a numerical demonstration can never be a substitute for a
rigorous mathematical proof, the present work provides the strongest
evidence to date that the instability cascade may exist in the case of
the temperature patch.  The existence of such a cascade would in turn
have implications for the regularity problem of the quasigeostrophic
equations when the initial $\theta$ is a smooth distribution in space,
a problem about which very little is known.  Because $\theta$ is
conserved on fluid particles, a collapsing filament with $\theta$
smoothly distributed across its width will retain its initial peak
$\theta$ value at the filament center.  As pointed out in
Ref.\cite{scott:2011} the filament boundary, or zero contour of the
$\theta$ field, may be identified with the patch contour described
above, in which case an instability cascade would again be expected to
lead to the unbounded growth in $\nabla\theta$ in finite time.  This
is in contrast to the case of the corner singularity of
Ref.\cite{cordoba:2005}, which is intrinsic to the case of the
temperature patch and which has no counterpart in the case of an
initially smooth $\theta$ distribution.  The present results suggest,
therefore, that the evolution of the initially smooth $\theta$
distribution may also develop a singularity in finite time, an open
problem of considerable theoretical interest.

\bibliography{./ssing}


\end{document}